\documentclass[11pt]{article}

\usepackage{graphicx, amsmath, amssymb, amsfonts, mathtools, mathrsfs, color}
\usepackage{comment, enumerate, tabularx, todonotes}
\usepackage{hyperref, url}
\usepackage[margin=1in]{geometry}

\newcommand{\pd}[2]    { \frac{\partial #1} {\partial #2} }

\newcommand{\pdi}[2] { {\partial_#2} #1 }
\newcommand{\td}[2] { \frac{d #1} { d #2 } }

\newcommand{\abs}[1]{\left| #1 \right|}

\newcommand{\mean}[1]{\left< #1 \right>}
\newcommand{\eps}{\varepsilon}
\newcommand{\dx}{\, dx}
\newcommand{\CC}{{\mathbb{C}}}

\newcommand{\freqp}{f_p}
\newcommand{\etastd}{\eta_{\text std}}
\newcommand{\depth}{d}
\newcommand{\dup}{\depth_{-}}
\newcommand{\ddn}{\depth_{+}}
\newcommand{\dupdn}{\depth_{\pm}}
\newcommand{\lam}{\lambda}
\newcommand{\lamup}{\lam_{-}}
\newcommand{\lamdn}{\lam_{+}}
\newcommand{\lamupdn}{\lam_{\pm}}
\newcommand{\lamfac}{N}
\newcommand{\drat}{\mathcal{D}}
\newcommand{\dratdn}{\drat_+}
\newcommand{\dratupdn}{\drat_{\pm}}
\newcommand{\omavg}{\omega_0}
\newcommand{\omsig}{\sigma_{\omega}}
\newcommand{\Dphi}{\Delta \phi}
\newcommand{\En}{\mathcal{E}}
\newcommand{\Mo}{\mathcal{M}}
\newcommand{\skw}{\text{skew}}
\newcommand{\skwdn}{\skw_+}
\newcommand{\var}{\text{var}}
\newcommand{\varup}{\var_-}
\newcommand{\vardn}{\var_+}
\newcommand{\varupdn}{\var_{\pm}}
\newcommand{\kurt}{\text{kurt}}
\newcommand{\std}{\text{std}}

\newcommand{\ampscale}{\mathcal{A}}
\newcommand{\lengthscale}{\mathcal{L}}
\newcommand{\timescale}{\mathcal{T}}
\newcommand{\epsup}{\eps_0}
\newcommand{\delup}{\delta_0}
\newcommand{\uhat}{\hat{u}}
\newcommand{\sympJ}{\mathcal{J}}
\newcommand{\vard}[2]{\frac{\delta #1}{\delta #2}}
\newcommand{\Ham}{\mathcal{H}}
\newcommand{\Hup}{\Ham^{-}}
\newcommand{\Hdn}{\Ham^{+}}
\newcommand{\Hupdn}{\Ham^{\pm}}
\newcommand{\Hthree}{\Ham_{3}}
\newcommand{\Htwo}{\Ham_{2}}
\newcommand{\Fcnl}{\mathcal{F}}
\newcommand{\Proj}{\mathcal{P}_{\Lambda}}
\newcommand{\uL}{u_{\Lambda}}
\newcommand{\HLupdn}{\Ham_{\Lambda}^{\pm}}
\newcommand{\SympL}{\sympJ_{\Lambda}}

\newcommand{\uupdn}{u_{\pm}}
\newcommand{\Gibbs}{\mathcal{G}}
\newcommand{\Gup}{\Gibbs^{-}}
\newcommand{\Gdn}{\Gibbs^{+}}
\newcommand{\Gupdn}{\Gibbs^{\pm}}
\newcommand{\thup}{\theta^{-}}
\newcommand{\thdn}{\theta^{+}}
\newcommand{\thupdn}{\theta^{\pm}}
\newcommand{\meanup}[1]{\mean{#1}_{-}}
\newcommand{\meandn}[1]{\mean{#1}_{+}}
\newcommand{\meanupdn}[1]{\mean{#1}_{\pm}}
\newcommand{\Gz}{\Gibbs_0}
\newcommand{\meanz}[1]{\mean{#1}_0}

\newcommand{\tavg}[1]{\overline{#1}}
\newcommand{\transf}{\mathcal{F}}
\newcommand{\Nsamp}{N_s}
\newcommand{\sumsamp}{\sum_{i=1}^{\Nsamp}}
\newcommand{\Fth}{W}
\newcommand{\RHS}{F}
\newcommand{\RHSh}{\hat{\RHS}}
\newcommand{\uhvec}{\mathbf{u}}
\newcommand{\RHSvec}{\mathbf{F}}
\newcommand{\usqdah}{\hat{v}} 

\newcommand{\boetal}{Bolles {\it et al.}~(2019)~}
\newcommand{\maetal}{Majda {\it et al.}~(2019)~}

\begin{document}

\title{Anomalous waves triggered by abrupt depth changes: laboratory experiments and truncated KdV statistical mechanics}

\author{
M.~N.~J.~Moore\thanks{Florida State University}, 
C.~Tyler Bolles\thanks{University of Michigan},
Andrew J.~Majda\thanks{Courant Institute of Mathematical Sciences}, 
Di Qi\footnotemark[3] }
\maketitle

\begin{abstract} 
Recent laboratory experiments of \boetal demonstrate that an abrupt change in bottom topography can trigger anomalous statistics in randomized surface waves. Motivated by these observations, \maetal developed a theoretical framework, based on deterministic and statistical analysis of the truncated Korteweg–de Vries (TKdV) system, that successfully captures key qualitative features of the experiments, including the robust emergence of anomalous statistics and heightened skewness in the outgoing wavefield. Here, we extend these parallel experimental and modeling efforts with several new findings that have resulted from a synergetic interaction between the two. By precisely relating model parameters to physical ones, we calibrate the model inverse temperature to the specific conditions present in the experiments, thereby permitting a quantitative comparison. We find theoretically predicted distributions of surface displacement to match the experimental measurements with surprising detail. Prompted by the presence of surface slope in the TKdV Hamiltonian, we present new experimental measurements on surface slope statistics and compare them to model predictions. Analysis of some deterministic trajectories of TKdV elucidates the experimental length and time scales required for the statistical transition to a skewed state. Finally, the theory predicts a peculiar relationship between the outgoing displacement skewness and the change in slope variance, specifically how their ratio depends on the wave amplitude and depth ratio. New experimental measurements confirm this prediction in spectacular fashion.
\end{abstract}

\section{Introduction}

Rogue waves are abnormally large surface waves, defined by oceanographers as those that exceed twice the significant wave height \cite{muller2005rogue, ying2011linear}. Though such waves were once dismissed as myth, they have now been recorded in oceans across the globe and pose a recognized threat to seagoing vessels and naval structures. Rogue waves, also variously known as freak or anomalous waves, have been observed in shallow \cite{pelinovsky2000nonlinear, gramstad2013freak}, intermediate \cite{karmpadakis2019laboratory}, and deep water \cite{dematteis2018rogue, dematteis2019experimental}, and certain rogue features have even been recovered by exact solutions to various wave models \cite{peregrine1983water, clarkson2017rational, chen2019periodic}.
These abnormal waves can be triggered by a variety of mechanisms, including anomalous wind-forcing \cite{kharif2008influence, toffoli2017wind}, opposing currents \cite{garrett2009rogue, onorato2011triggering}, focusing due to variable bathymetry \cite{heller2008refraction, white1998chance}, 
and the Benjamin-Feir deep-water modulational instability \cite{benjamin1967disintegration, viotti2013emergence, cousins2015unsteady, farazmand2017reduced}.
The common tie between these mechanisms is their ability to generate non-normal statistics in the surface displacement: when governed by Gaussian statistics, the likelihood of a rogue wave is extremely low, but these events occur much more frequently when surface statistics deviate from Gaussian. In this way, anomalous waves can be approached from the broader perspective of turbulent dynamical systems \cite{sapsis2013a, sapsis2013b, sapsis2013blending, chen2016filtering, majda2016introduction, macedo2017universality, MajdaQiSIAM2018, blonigan2019extreme, guth2019machine, holm2019stochastic}.

A recent series of laboratory \cite{bolles2019, trulsen2020extreme} and numerical investigations \cite{viotti2014, herterich2019extreme} have demonstrated the emergence of anomalous wave statistics from abrupt variations in bottom topography. Since topographical variations are strictly one-dimensional, these studies can be viewed as offering a bare minimum set of conditions capable of generating anomalous waves. In particular, the more complex mechanism of focusing by 2D bottom topography is  absent. The studies thus offer a paradigm system, with emergent anomalous features similar to those seen in more complex systems, but in a tractable context that is amenable to analysis.

Our particular focus is the laboratory experiments of \boetal~\cite{bolles2019}, who demonstrated the emergence of anomalous statistics from a randomized wave-field encountering an abrupt depth change (ADC). In these experiments, the incoming wave-field is generated with nearly Gaussian statistics, and a plexiglass step placed near the middle of the tank creates the depth change. Upon passing over the step, the wave distribution skews strongly towards positive displacement, with deviation from Gaussian being most pronounced a short distance downstream of the ADC. Inspired by these experimental observations, \maetal \cite{majda2019} developed a theoretical framework that accurately captures several key aspects of the anomalous behavior. The theory is based on deterministic and statistical analysis of the truncated Korteweg–de Vries (TKdV) equations, and uses a combination of computational, statistical, and analytical tools. In subsequent work, Majda \& Qi (2019) \cite{majdaqi2019} analyzed more severely truncated systems --- as low as two interacting modes --- that exhibit a statistical phase transition to anomalous statistics and the creation of extreme events while enjoying a more tractable structure. More recently, Qi \& Majda (2019) \cite{MachineLearning2019} demonstrated the capability of machine learning strategies to accurately predict these extreme events. That work employs a deep neural network coupled with a judicious choice of the entropy loss function that emphasizes the dominant structures of the turbulent field \cite{MachineLearning2019}. It should be noted that, in these studies, extreme events are represented by strong skewness of the wave-field. They occur intermittently and on a relatively frequent basis, which is in contrast to the simpler situation of isolated, rare events \cite{guth2019machine}.

The purpose of the present manuscript is to provide a more comprehensive treatment of the {\em combined} experimental and theoretical efforts of \boetal~\cite{bolles2019} and \maetal~\cite{majda2019, majdaqi2019}. We present a variety of new findings that have resulted from a synergetic interaction between theory and experiments -- a cooperative strategy with a proven record of success \cite{camassa2012stratified, ristroph2012, ganedi2018equilibrium}. 
That outline of the paper is as follows. In Section \ref{experiments} we detail the laboratory experiments of \boetal \cite{bolles2019} and summarize previously reported findings on anomalous statistics of the surface displacement. We also present new experimental data on surface {\em slope} statistics --- a line of inquiry motivated by the theoretical advancements of \maetal \cite{majda2019}, specifically the central role played by the slope in the Hamiltonian structure of TKdV.  Section \ref{theory} discusses the TKdV theoretical framework, including both the deterministic and the statistical mechanics perspectives. In this section, we flesh out details of the non-dimensionalization, which, for brevity, were only briefly discussed in previous work. This exercise provides a more precise link between model and experimental parameters, ultimately facilitating a closer comparison between the two. We then discuss TKdV as a deterministic dynamical system, approachable from the viewpoint of statistical mechanics. The later perspective is based on the Hamiltonian structure and novel ensemble distributions that incorporate canonical and micro-canonical aspects \cite{abramov2003}. 

Section \ref{results} presents a systematic comparison between experiments and theory, focusing on both surface displacement and surface slope. We find the statistical distributions of these two quantities to agree well across experiments and theory. In particular, we find remarkably quantitative agreement in the displacement statistics. We then examine numerical simulations of the TKdV deterministic dynamics, which elucidates the length and timescales required for statistical transitions in the experiments. Finally, this section discusses a peculiar power-law relationship between surface-displacement skewness and surface-slope variance predicted by the statistical mechanics framework \cite{majda2019}. New experimental measurements conclusively confirm this prediction, further demonstrating the predictive power of the TKdV framework developed by \maetal \cite{majda2019}. We close with some final remarks in Section \ref{conclusion}.

\section{The experiments}
\label{experiments}

\begin{figure}
\begin{center}
\includegraphics[width = 0.80 \linewidth]{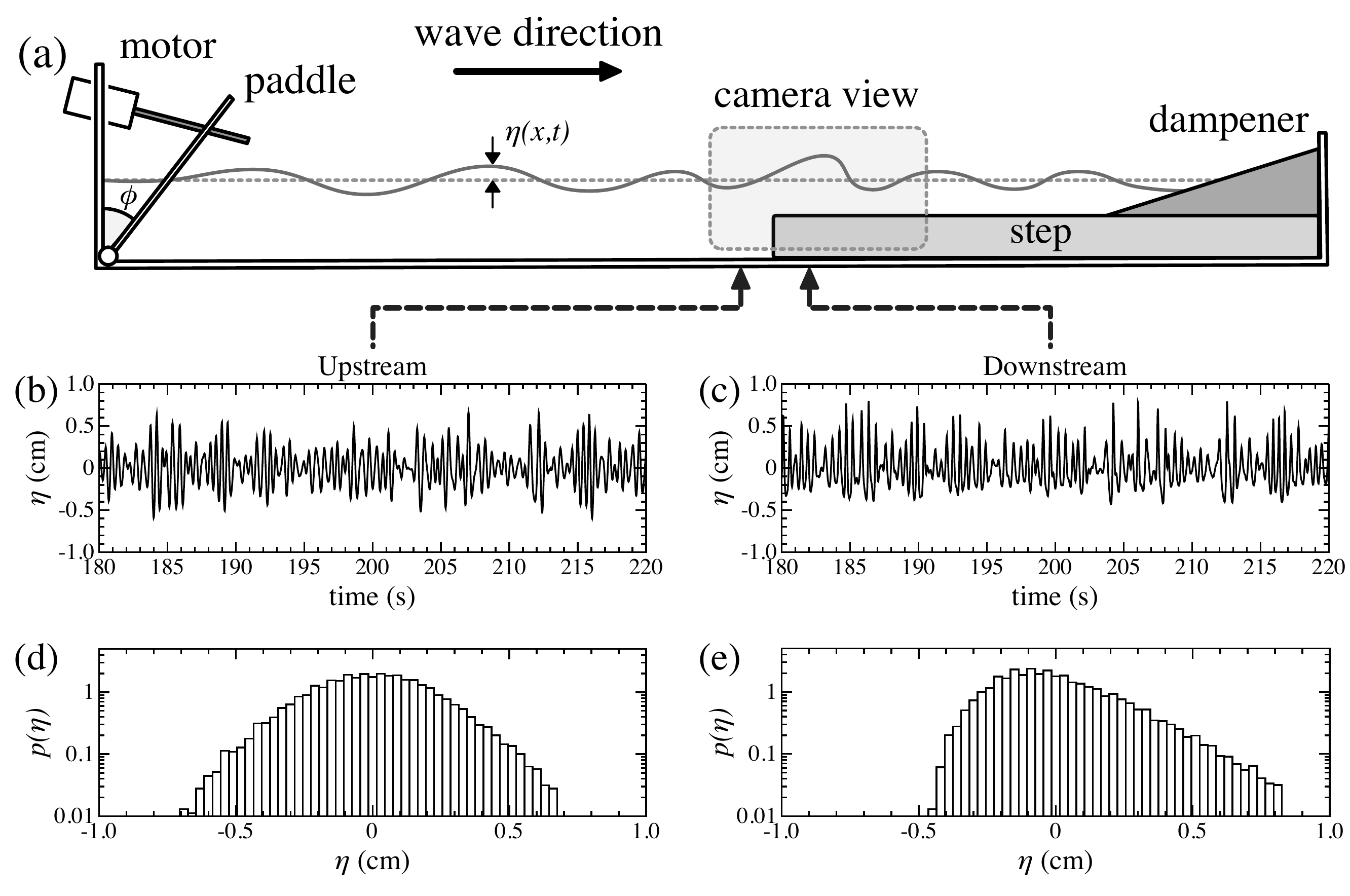}
\caption{
(a) Experimental schematic: randomized waves are generated by a pivoting paddle and propagate over a step in bottom topography.
(b)--(c) Surface displacement measured at representative locations upstream and downstream of the abrupt depth change (ADC). (d)--(e) Corresponding histograms showing symmetric upstream statistics and highly skewed downstream statistics. Figure adapted from \cite{bolles2019}.}
\label{ExpDiagStats}
\end{center}
\end{figure}

	As diagrammed in Fig.~\ref{ExpDiagStats}(a), the experiments consist of a long, narrow wave tank (6 m long x 20 cm wide x 30 cm high), with waves generated by a plexiglass paddle at one end \cite{bolles2019}. The waves propagate through the tank and, roughly midway through, pass over an abrupt depth change (ADC) created by a plexiglass step. The waves continue to propagate through the shallower depth until reaching the far end of the tank, at which point their energy is dissipated by a horse-hair dampener. Since the dampener minimizes backscatter, the waves in this experiment propagate primarily in one direction, from left to right in Fig.~\ref{ExpDiagStats}(a). 

The pivoting motion of the paddle is driven by a 5-phase stepper motor. To generate a randomized wave field, the paddle angle $\phi$ is specified by a psuedo-random signal
\begin{align}
\label{PaddleAngle}
& \phi(t) = \phi_0 + \Dphi \sum_{n=1}^N a_n \cos(\omega_n t+\delta_n) \, , \\
\label{anEq}
& a_n = \sqrt{\frac{2 \Delta \omega}{\pi^{1/2} \omsig}} \, 
\exp \left( -\frac{(\omega_n - \omavg)^2}{2 \omsig^2} \right) \, ,
\end{align}
The angular frequencies are evenly spaced $\omega_n = n  \Delta \omega$ with step size $ \Delta \omega = (\omavg+4 \omsig)/N$, where $\omavg$ and $\omsig$ represent the mean and the bandwidth of $\omega$ respectively. As in prior work, all experiments reported here use the values $\omavg = \omsig = 12.5$ rad/s, corresponding to a peak forcing frequency of 2 Hz and bandwidth of 2 Hz. The phases $\delta_n$ are uniformly distributed random variables, which results in a randomized wave train. The standard deviation of the paddle angle, $\Dphi$, controls the overall amplitude of the waves. In a single experiment $\Dphi$ is fixed, and we will present a series of experiments with $\Dphi$ varied systematically.

The free surface is illuminated by light-emitting diodes and is imaged from the sideview with a Nikon D3300 at 60 frames per second. The illumination technique, coupled with high pixel count of the camera, allows surface displacements to be resolved with accuracy better than 1/3 millimeter. Furthermore, these optical measurements permit extraction of wave statistics {\em continuously} in space, rather than at a few discrete locations, which is crucial for identifying regions of anomalous wave activity. Further details of the experimental setup can be found in \cite{bolles2019}.
	
Example measurements of free-surface displacements $\eta$ are shown in Figs.~\ref{ExpDiagStats}(b)--(c). These measurements are extracted from the images at two representative locations: one a short distance (9 cm) upstream of the ADC and the other a short distance (15 cm) downstream. Both signals exhibit a combination of periodic and random behavior, with the dominant oscillations corresponding to the peak forcing frequency of 2 Hz. 
	
The nature of the random fluctuations is revealed by the corresponding histograms shown in Fig.~\ref{ExpDiagStats}(d)--(e) on a semi-log scale. The upstream measurements are symmetrically distributed about the mean, $\eta = 0$. In fact, \boetal found that these measurements follow a Gaussian distribution closely \cite{bolles2019}. The downstream measurements, however, skew strongly towards positive displacement, $\eta > 0$. \boetal found these measurements to be well described by a mean-zero gamma distribution \cite{bolles2019}. The slower decay of the gamma distribution indicates an elevated level of extreme surface displacement, i.e.~rogue waves. \boetal estimated that a rogue wave can be up to 65 times more likely in these experiments than if displacements were Gaussian \cite{bolles2019}. 

\begin{figure}
\begin{center}
\includegraphics[width = 0.8 \linewidth]{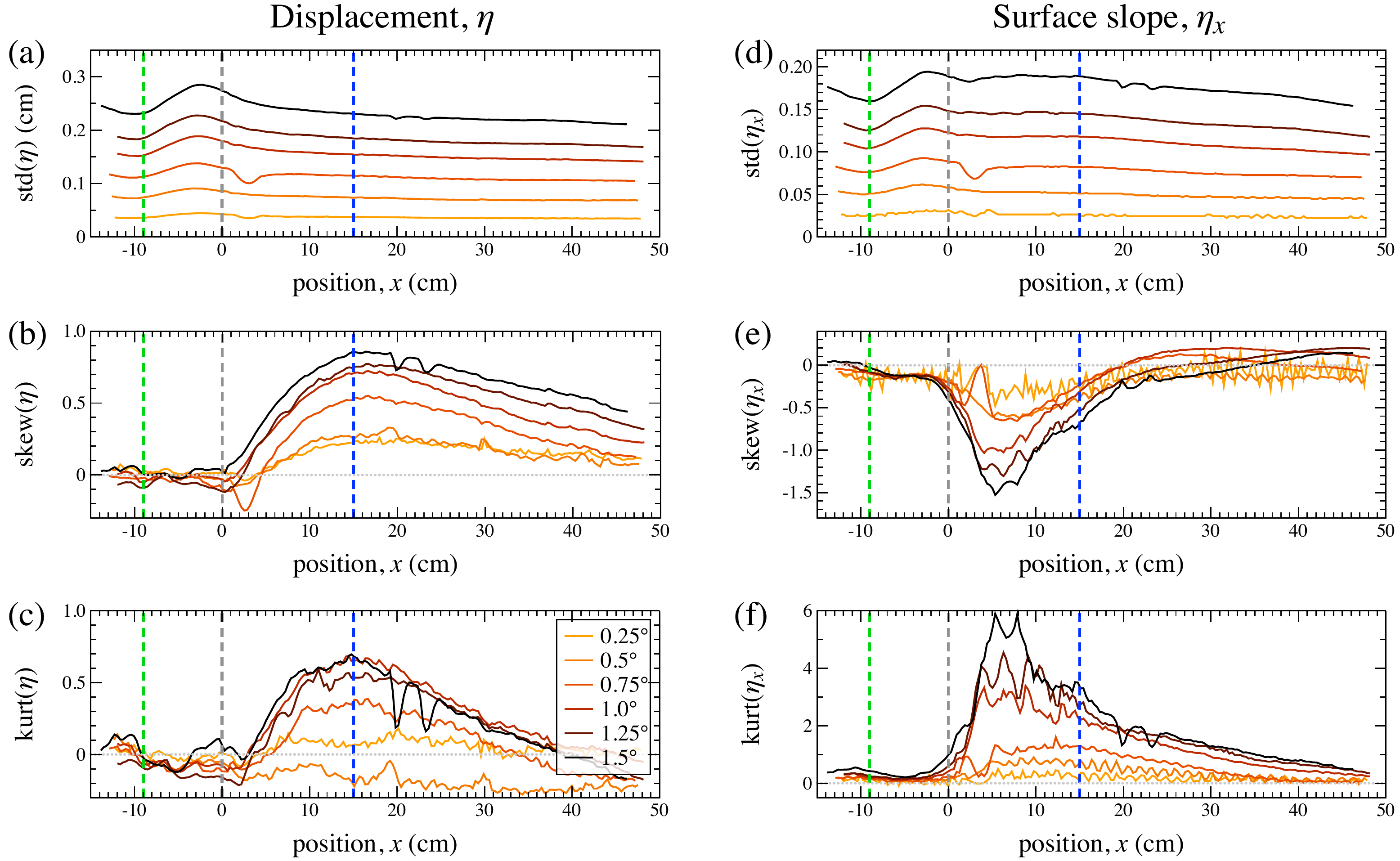}
\caption{Wave statistics as they vary in space for several experiments of different driving amplitudes (see legend). (a)--(c) Standard deviation, skewness, and kurtosis of the surface displacement $\eta$. (d)--(f) The same for surface slope $\eta_x$. (a) $\etastd$, which sets a scale for wave amplitude, does not vary significantly crossing the ADC. (b)--(c) Skewness and kurtosis of $\eta$, however, show a strong response to the ADC. (d)--(f) The measurements of surface slope, $\eta_x$, are negatively skewed and also exhibit large kurtosis downstream of the ADC. }
\label{ExpSpatialStats}
\end{center}
\end{figure}

The paddle amplitude in Fig.~\ref{ExpDiagStats} is $\Dphi = 1.38^{\circ}$, and this value was varied systematically in the range $\Dphi = 0.125^{\circ}$--$2^{\circ}$ to probe the different regimes of wave behavior, from linear to strongly nonlinear waves. Figure \ref{ExpSpatialStats} shows long-time statistics of both surface displacement $\eta$ and the surface slope $\eta_x$ as they vary in space for six different driving amplitudes (see legend). 

In this paper, we only examine statistic of mean-zero quantities, and so, for an arbitrary mean-zero quantity $q$, we have the following definitions
\begin{align}
& \std(q) = q_{std}= \sqrt{\mean{q^2}}
&&\mbox{\em standard deviation} \\
& \skw(q) = {\mean{q^3}} / {q_{std}^3}	
&&\mbox{\em skewness} \\
& \kurt(q) = {\mean{q^4}} / {q_{std}^4} - 3
&&\mbox{\em (excess) kurtosis}
\end{align}
where $\mean{}$ indicates a mean --- here a long-time mean at a fixed spatial location. Hereafter, we will simply refer to the excess kurtosis as kurtosis.
  
	Figure \ref{ExpSpatialStats} shows how these statistics vary in the vicinity of the ADC, located at $x = 0$, for both displacement and slope. First, the standard deviation of displacement, $\etastd$, gives the coarsest possible estimate for the amplitude of waves. Figure \ref{ExpSpatialStats}(a) shows that while $\etastd$ increases with driving amplitude, it remains relatively uniform in space for each individual experiment, indicating that the overall amplitude of the wave train is not significantly altered by the presence of the ADC. The skewness and kurtosis, however, respond strongly to the ADC as long as the amplitude is sufficiently high. As seen in \ref{ExpSpatialStats}(b)--(c), both the skewness and kurtosis are relatively small upstream of the ADC, indicating nearly Gaussian statistics, but then increase dramatically downstream and reach a peak near $x = 15$ cm. Interestingly, the location of the peak is the same for skewness and kurtosis and appears insensitive to driving amplitude. The maximum values of skewness and kurtosis (roughly 0.9 and 0.7 respectively) seen in the figure indicate a significant departure from Gaussian statistics, as is constant with the histogram in Fig.~\ref{ExpDiagStats}(e). \boetal found that, once a threshold driving amplitude is exceeded (roughly $\Dphi = 0.5^{\circ}$), the displacement statistics in this anomalous region are robustly described by the gamma distribution across all of the experiments. We note that we have gathered data for 15 different driving amplitudes, but only display 6 in Fig.~\ref{ExpSpatialStats} to avoid clutter. 
 
	To complement the displacement statistics, we report here statistics for the surface slope $\eta_x$ in the right column of Fig.~\ref{ExpSpatialStats}. Our attention to slope statistics was motivated by new theoretical developments by \maetal \cite{majda2019}, as will be expanded upon in later sections. We extract the surface slope by numerical differentiating images of the free surface using Savitzky-Golay smoothing filters. This ability to extract surface slope is another advantage of our optical measurements over the more commonly used technique of placing a set of discrete wave probes in the tank. As seen in Fig.~\ref{ExpSpatialStats}(a) the standard deviation of slope behaves similar to displacement: $\std(\eta_x)$ increases with driving amplitude, but remains nearly uniform in space for each individual experiment and is not affected by the ADC. The higher-order moments, however, respond strongly to the ADC if the driving amplitude is sufficiently high. The skewness of slope becomes highly negative and reaches a minimum near $x = 8$ cm, which is downstream of the ADC but upstream of the position where the displacement statistics peak. The kurtosis of $\eta_x$ reaches a peak at nearly the same location as skewness. We note that the large {\em negative} skewness of $\eta_x$ indicates a bias towards negative slope, which is consistent with a right-moving wave of steep leading surface and shallower trailing surface, i.e.~a wave that is near overturning.

To have a simple baseline for comparison against theory, we refer to the data from Fig.~\ref{ExpDiagStats} as the {\em reference} experiments. In these experiments, the driving amplitude is $\Dphi = 1.38^{\circ}$ (intermediate between the two largest amplitudes shown in Fig.~\ref{ExpSpatialStats}), the upstream depth is $\dup = 12.5$ cm, and the downstream depth is $\ddn = 3$ cm, giving a depth ratio of $\dratdn = \ddn/\dup= 0.24$. From the data represented in Fig.~\ref{ExpSpatialStats}, we extract a characteristic wave amplitude for the representative experiments of $\etastd = 0.21$ cm, which will be important for setting dimensionless parameters that enter the theory. Table \ref{paramtable} lists the range of experimental parameters and the representative values, as well as values of dimensionless parameters that will be introduced later.

\begin{table}[h]
\begin{center}
\caption{Table of parameters}
\label{paramtable}
\begin{tabular}{l l l l}
\hline Description & Notation & Experimental range & `Representative' values \\
\hline
Peak forcing frequency	& $f_p$					& 2 Hz 		& 2 Hz \\
Characteristic amplitude	& $\etastd$				& 0.03--0.3 cm	& 0.21 cm \\
Upstream depth		& $\dup$					& 12.5 cm 	& 12.5 cm \\
Downstream depth		& $\ddn$					& 2.2--5.3 cm	& 3 cm \\
Upstream wavelength	& $\lamup = \sqrt{g \dup}/f_p$	& 55 cm 		& 55 cm \\
Downstream wavelength	& $\lamdn = \sqrt{g \ddn}/f_p$	& 23--36 cm	& 27 cm \\
Amplitude-to-depth ratio	& $\epsup = \etastd / \dup$	&0.0024--0.024	& 0.017 \\
Depth-to-wavelength ratio	& $\delup = \dup / \lamup$	& 0.23		& 0.23 \\
Depth ratio			& $\dratdn = \ddn/\dup$		& 0.18--0.42	& 0.24
\end{tabular}
\end{center}
\end{table}

\section{Theoretical framework}
\label{theory}

We now introduce the theoretical framework that will be used to understand and quantify the experimental observations. This framework is based on a Galerkin truncation of the variable-depth Korteweg–de Vries (KdV) equation. The KdV equation is a well established model for describing the propagation of unidirectional, shallow-water waves, accounting for weak nonlinearity and weak dispersion over long timescales and large spatial scales. We will perform Galerkin truncation of KdV to obtain a finite-dimensional dynamical system that exhibits weak turbulence. We outline the Hamiltonian structure of both the traditional KdV and the truncated systems. This structure is exploited to obtain invariant measures of the underlying dynamics and, ultimately, to rationalize the experimental findings on anomalous wave statistics triggered by an ADC.

\subsection{The Korteweg–de Vries equation with variable depth}

We consider the surface displacement $\eta(x,t)$ of unidirectional, shallow-water waves in a reference frame moving with the characteristic wave speed, $\xi = x - ct$. Here, $c = \sqrt{g \depth}$ is leading-order approximation to the wave speed (i.e.~from linear theory), where $g$ is gravity and $\depth$ is the local depth.
The leading-order dynamics are corrected to first order in small amplitude by the Korteweg–de Vries equation (KdV), which in dimensional form is given by \cite{whitham2011linear}
\begin{equation}
\label{KdV}
\eta_t + \frac{3 c}{2 \depth} \eta \eta_{\xi} + \frac{c \depth^2}{6} \eta_{\xi \xi \xi} = 0
\end{equation}
Motivated by the experiments, we consider waves that originate from a region of constant depth, encounter an abrupt depth change, and continue into another region of constant depth. Thus, depth will be piecewise constant
\begin{align}
\depth = 
\begin{cases}
\dup \quad \mbox{if } x<0 \\
\ddn \quad \mbox{if } x>0
\end{cases}
\end{align}
Most often, we consider waves moving into shallower depth, so that $\dup > \ddn$. Throughout this paper, we use the subscript `-' to represent upstream variables and `+' for downstream variables.

In the experiments, the randomized incoming wave-field is generated with a peak forcing frequency of $\freqp = 2$ Hz, which gives rise to the characteristic wavelength of $\lam = c/\freqp = \sqrt{g \depth} / \freqp$. Note that both the characteristic wave speed $c = c_{\pm}$ and wavelength $\lam = \lamupdn$ take different values upstream and downstream of the ADC. We remark that experimental measurements indicate that $\etastd$ is nearly the same on both sides of the ADC. Hence, we will not distinguish between upstream and downstream values of $\etastd$.

\subsection{Nondimensionalization and relation to experimental scales}

In this section, the variable-depth KdV equation \eqref{KdV} will be recast into a dimensionless form that is chosen for convenience in working with the statistical-mechanics framework of \maetal \cite{majda2019}. Since the choice of normalization is not unique, it is instructive to first introduce a generic normalization to facilitate comparison with other possible choices. To this end, we consider characteristic scales, $\ampscale, \lengthscale, \timescale$ for the wave amplitude, longitudinal length, and time respectively, which can remain unspecified for the moment. We introduce the dimensionless variables
\begin{align}
&u = \eta / \ampscale
&&\mbox{\em dimensionless surface displacement} \\
&\tilde{x} = (x-ct) / \lengthscale
&&\mbox{\em dimensionless position (in moving frame)} \\
&\tilde{t} = t / \timescale
&&\mbox{\em dimensionless time}
\end{align}
Recasting \eqref{KdV} in terms of these variables gives the generic dimensionless KdV equation:
\begin{equation}
u_t + \frac{3}{2} \left( \frac{c \timescale \ampscale}{\lengthscale \depth} \right) u u_x 
+ \frac{1}{6} \left( \frac{c \timescale \depth^2}{\lengthscale^3} \right) u_{xxx} = 0
\end{equation}
We have dropped the tilde notation above for simplicity and will henceforth use tildes only in cases of possible ambiguity.

Now it is possible to choose the scales $\ampscale, \lengthscale, \timescale$ for ease in working with a particular framework. We make the following choices,
\begin{align}
\label{scales}
\ampscale = \pi^{1/2} \, \etastd \, , \qquad
\lengthscale_{\pm} = \frac{\lamfac \lamupdn}{2 \pi} \, , \qquad
\timescale_{\pm} = \frac{\lamfac \lamupdn}{2 \pi \freqp \dupdn}
\end{align}
where $\lamfac$ is an integer to be chosen later. 
The explanation for these choices is as follows. First, we have chosen the characteristic amplitude, $\ampscale$, to normalize the energy of the state-variable $u$ to unity, as will be demonstrated in Section \ref{tKdVSec}. Second, regarding $\lengthscale$, recall that $\lam$ is the characteristic wavelength corresponding to the peak forcing frequency $\freqp$ in the experiments. If only integer multiples of $\freqp$ were imposed (e.g.~lower frequencies were not present), then the forcing would produce waves that are periodic over lengthscale $\lam$. Since lower frequencies do exist, strict periodicity is not satisfied, but rather waves may be nearly periodic over the physical domain $\xi \in [-\lam/2, \lam/2]$. The approximation of near-periodicity becomes more accurate if integer multiples are considered, i.e.~$\xi \in [-\lamfac \lam/2, \lamfac \lam/2]$. Thus, we have chosen $\lengthscale$ above so that, over the dimensionless domain $\tilde{x} \in [-\pi, \pi]$, periodic boundary conditions can be  imposed on $u$ with an accuracy that increases with $\lamfac$. 

	Lastly, regarding the characteristic timescale $\timescale$, the most basic timescale in the experiments is simply $\freqp^{-1}$, i.e.~the period of waves passing a fixed reference point. Of course, the leading-order behavior in shallow water is simply wave propagation with uniform speed $c$, i.e.~no dispersion. The KdV equation provides the first correction to this behavior and describes dynamics that evolve over longer timescales. Hence we have rescaled $\freqp^{-1}$ by the factor $N \lam/(2 \pi \depth) \gg 1$, which provides a suitably long timescale in line with other normalizations \cite{johnson1997modern}. The scales $\lengthscale = \lengthscale_{\pm}$ and $\timescale = \timescale_{\pm}$ change value across the ADC, which is important to note when comparing the theory against experimental measurements.

With the above choices, the dimensionless KdV equation takes the form
\begin{align}
\label{dimlessKdV}
&u_t + C_3 \drat^{-3/2} \, u u_x + C_2 \drat^{1/2} \, u_{xxx} = 0
\qquad \text{for } x \in [-\pi,\pi] \\
\label{C3C2}
&C_3 = \frac{3}{2} \pi^{1/2} \epsup \delup^{-1} \, , \quad
C_2 = \frac{2 \pi^2 \delup}{3 \lamfac^2} 
\end{align}
The constants $C_3$ and $C_2$ do {\em not} change value crossing the ADC and are given in terms of the dimensionless parameters
\begin{align}
&\epsup = \etastd / \dup
&&\mbox{\em upstream amplitude-to-depth ratio} \\
&\delup = \dup / \lamup
&&\mbox{\em upstream depth-to-wavelength ratio}
\end{align}
The reason for the subscripts $3$ and $2$ will become evident in the next section. 

Meanwhile, the dimensionless depth $\drat = {\depth}/{\dup}$ {\em does} change value across the ADC since the depth $\depth$ changes. Recall that the reference frame of \eqref{dimlessKdV} moves with the local wave speed via the variable $\xi = x-ct$ from \eqref{KdV}. Thus, the ADC is met at some time $T_{ADC}$, and for simplicity we set $T_{ADC} = 0$. Therefore, we can regard $\drat$ as a piece-wise-constant function of dimensionless time
\begin{equation}
\label{dratpw}
\drat = 
\begin{cases}
1 		&\quad \mbox{for } {t}<0 \\
\dratdn = {\ddn}/{\dup} 	&\quad \mbox{for } {t}>0
\end{cases}
\end{equation}
See Table \ref{paramtable} for a summary of these dimensionless parameters and their values in experiments.

A few comments are in order. First, we note that the original formulation of this theory utilized a slightly different normalization \cite{majda2019}, with identical powers of $\drat$ in \eqref{dimlessKdV} but with different expressions for the other dimensionless parameters. These differences are purely cosmetic, and we have made the choices above simply to facilitate comparison with experiments. Second, an alternate formulation of the variable-depth KdV equation has been proposed in which the product $\depth^{1/4} \eta$, rather than $\eta$, is conjectured to vary continuously across the ADC \cite{johnson1997modern}. Of course, that assumption implies a discontinuity in surface displacement, which, though perhaps small, would be physically unrealistic. We have chosen to enforce continuity of surface displacement on the basis of physical realism. Furthermore, our {\em direct experimental measurements} of $\etastd$ give no indication of a significant change across the ADC, thus supporting the formulation used here. We note, however, that the only modification resulting from the alternate formulation would be in the power of $\drat$ in the second term of \eqref{dimlessKdV}: the power $\drat^{-3/2}$ would become $\drat^{-7/4}$. Thus, in this alternate formulation, the second term in \eqref{dimlessKdV} would be scaled by a negative power of $\drat$ and the third term by exactly the same positive power of $\drat$. Hence, the two formulations are qualitatively very similar with only slight quantitative differences expected.

\subsection{Hamiltonian structure of KdV}
\label{HamiltonianSection}

The variable-depth KdV \eqref{dimlessKdV}, though not Hamiltonian throughout the entire domain, admits a Hamiltonian structure on each side of the ADC. Indeed, \eqref{dimlessKdV} can be expressed as
\begin{align}
\label{HamStruct}
\partial_t{u} = \sympJ \vard{\Hupdn}{u}
\end{align}
where $\sympJ = \pdi{}{x}$ is the symplectic operator and $\Ham = \Hupdn$ is the Hamiltonian, which takes different forms on either side of the ADC. It is convenient to decompose the Hamiltonian into a so-called cubic and quadratic component, given respectively by
\begin{align}
\label{H3H2}
\Hthree = \frac{1}{6} \int_{-\pi}^{\pi} u^3 \dx	\, , \qquad
\Htwo = \frac{1}{2} \int_{-\pi}^{\pi} u_x^2 \dx	\, .
\end{align}
Then the Hamiltonian can be expressed as
\begin{equation}
\label{Hamiltonian}
\Hupdn = C_2 \dratupdn^{1/2} \, \Htwo - C_3 \dratupdn^{-3/2} \, \Hthree
\end{equation}
where $\drat = \dratupdn$ changes value across the ADC. More explicitly, substituting \eqref{dratpw} gives the separate upstream and downstream Hamiltonians as
\begin{align}
&\Hup = C_2 \, \Htwo - C_3 \, \Hthree 						&& \text{for } t<0 \\
&\Hdn = C_2 \dratdn^{1/2} \, \Htwo - C_3 \dratdn^{-3/2} \, \Hthree	&& \text{for } t>0
\end{align}
As seen in \eqref{H3H2}, the cubic component, $\Hthree$, represents the skewness of the wave-field, while the quadratic component, $\Htwo$, represents the energy of the surface slope. The sign difference between the two in \eqref{Hamiltonian} thus represents a competition between wave skewness and slope energy. In particular, the appearance of the slope energy in the theory motivated the new experimental measurements on surface {\em slope} statistics reported in this paper. 

We remark that, in defining the Hamiltonian, we have chosen the sign convention of Lax (1975) \cite{lax1975periodic}. More recent work of Bajars {\it et al.}~(2013) \cite{bajars2013weakly} and \maetal \cite{majda2019} use a different convention, in which both the signs of $\sympJ$ and $\Ham$ are opposite. Clearly, these sign differences cancel in \eqref{HamStruct} and thus the two conventions are completely equivalent. Using the second convention, \maetal found that a {\em negative} inverse temperature is required to accurately describe the experimental observations \cite{majda2019}. We have  chosen the convention above so that a {\em positive} inverse temperature may be used, allowing our theory to fit into the most standard form of statistical mechanics.

We introduce two important invariants of KdV, namely the momentum and the energy
\begin{align}
\label{MomEn}
\Mo[u] \equiv \int_{-\pi}^{\pi} u \dx \, = 0 , \qquad
\En[u] \equiv \frac{1}{2} \int_{-\pi}^{\pi} u^2 \dx = 1
\end{align}
As indicated above, the momentum of $u$ vanishes since it is measured as displacement from equilibrium. Second, due to the choice of $\ampscale$ in \eqref{scales}, the energy has been normalized to unity.

\subsection{Truncated KdV}
\label{tKdVSec}

We now introduce the truncated KdV (TKdV) system, which is the main focus of the present study. Consider the state variable represented as a spatial Fourier series
\begin{align}
&u(x,t) = \sum_{k=-\infty}^{\infty} \uhat_k(t) \, e^{i k x} \, , \\
\label{uhat}
&\uhat_k(t) = \frac{1}{2 \pi} \int_{-\pi}^{\pi} u(x,t) \, e^{-i k x} \dx \, ,
\end{align}
where $\uhat_k(t) \in \mathbb{C}$. Since $u(x,t)$ is real valued, $\uhat_{-k} = \uhat_{k}^*$, and since momentum vanishes $\uhat_0 = 0$.
Next, consider the Galerkin truncation at wave number $\Lambda$
\begin{align}
\uL(x,t) = \Proj u =
\sum_{\abs{k} \le \Lambda} \uhat_k(t) \, e^{i k x} \, , \qquad
\end{align}
where $\Proj$ is a projection operator and \eqref{uhat} still holds. Inserting the projected variable, $\uL$, into the KdV equation and applying the projection operator, $\Proj$, again where necessary produces the truncated KdV equation (TKdV)
\begin{align}
\label{TKdV}
&\pd{\uL}{t} +  \frac{1}{2} C_3 \drat^{-3/2} \, \pd{}{x} \Proj (\uL)^2 + C_2 \drat^{1/2} \, \frac{\partial^3 \uL}{\partial x^3} = 0
\qquad \text{for } x \in [-\pi,\pi] \\
&C_3 = \frac{3}{2} \pi^{1/2} \epsup \delup^{-1} \, , \quad C_2 = \frac{2 \pi^2 \delup}{3 \lamfac^2}
\end{align}
Note the additional projection operator in front of the quadratic term $\uL^2$, which removes the aliased modes of wavenumber larger than $\Lambda$. Since all wavenumbers larger than $\Lambda$ have been removed, \eqref{TKdV} represents a {\em finite} dimensional dynamical system, of dimension $\Lambda$ over $\mathbb{C}$. The constants $C_3$ and $C_2$ are the same as before and have been repeated here for convenience. 

Briefly, consider the parameter $\lamfac$, the number of characteristic wavelengths in the physical domain. We require $1 \le \lamfac \le \Lambda$, so that the mode $\uhat_{\lamfac}$, corresponding to the characteristic wavelength $\lam$ in the experiments, is resolved in the truncated dynamical system. If $\lamfac = \Lambda$, then $\lam$ corresponds to the smallest resolved wavelength. If instead $\lamfac$ is chosen as an intermediate value between 1 and $\Lambda$, then the truncated system will resolve scales that are both bigger and smaller than the characteristic value $\lam$.

Remarkably, the TKdV system \eqref{TKdV} retains the Hamiltonian structure described in Section \ref{HamiltonianSection}, with the only modification being the inclusion of the projection operator \cite{bajars2013weakly, majda2019}. The piecewise defined Hamiltonian for TKdV is given by
\begin{equation}
\label{TruncHamiltonian}
\HLupdn = C_2 \dratupdn^{1/2} \, \Htwo[\uL] - C_3 \dratupdn^{-3/2} \, \Hthree[\uL]
\end{equation}
where $\Hthree$ and $\Htwo$ are defined exactly as before \eqref{H3H2}, but now are simply applied to the projected variable $\uL = \Proj u$.
Then TKdV \eqref{TKdV} can be expressed as
\begin{align}
\partial_t {\uL} = \partial_x \Proj \, \vard{\HLupdn}{\uL}
\end{align}
where the truncated symplectic operator is $\SympL = \partial_x \Proj$.

The momentum and energy defined in \eqref{MomEn} remain invariants of TKdV, with the same normalized values $\Mo[\uL] = 0$ and $\En[\uL] = 1$. Note that Parseval's identity implies
\begin{equation}
\En[\uL] = 2 \pi \sum_{k=1}^{\Lambda} \abs{\uhat_k}^2 = 1
\end{equation}
Thus, the dynamics of interest are confined to the unit hypersphere, $\En = 1$ in $\CC^{\Lambda}$.
In summary, the TKdV system possesses three important invariants: momentum, energy, and Hamiltonian. The untruncated KdV equation possesses an infinite sequence of additional invariants \cite{lax1975periodic, whitham2011linear}, but their truncated counterparts are not generally invariants of TKdV.

\subsection{Mixed microcanonical-canonical Gibbs ensemble}

	In examining statistical mechanics of this Hamiltonian system, we will appeal to the idea of a {\em mixed microcanonical-canonical} Gibbs ensemble, as originally introduced by Abramov {\it et al.}~(2003) for the Burgers-Hopf system \cite{abramov2003}. Specifically, this ensemble is microcanonical in energy and canonical in the Hamiltonian. The reason this ensemble is needed is the sign indefiniteness of the cubic term $\Hthree$ in the Hamiltonian, which would cause a simple canonical distribution to diverge at infinity. The mixed ensemble, however, fixes the energy and hence confines dynamics to the compact set of the unit hypersphere $\En = 1$. Since the Hamiltonian is continuous, its value is bounded on the unit hypersphere, and thus the mixed ensemble produces a normalizable distribution. This construction applies equally well to the truncated or untruncated KdV system and hence we will not distinguish between the two. We note that other possible constructions may be applicable too \cite{kleeman2014nonequilibrium}.
	
	On either side of the ADC, the mixed ensemble, or {\em Gibbs measure}, follows directly from the corresponding Hamiltonian via
\begin{align}
\label{Gibbs}
\Gupdn = Z_{\thupdn}^{-1} \, \exp(-\thupdn \Hupdn) \delta(\En - 1)
\end{align}
Here $\theta = \thupdn$ is the inverse temperature, which will take a different value on either side of the ADC, and $Z_{\theta}$ a constant that depends on $\theta$. Each measure $\Gupdn$ induces a corresponding ensemble average, denoted $\mean{\cdot}_{\pm}$. 

	We note that, in the current formulation, a positive inverse temperature $\theta > 0$ produces physically realistic statistics with a decaying energy spectrum, as is consistent with experiments \cite{majda2019, bajars2013weakly}. Negative inverse temperature produces a physically unrealistic spectrum that has more energy at smaller scales. Hence, we will hereafter focus on the physically realistic case of $\theta > 0$.

\subsection{Matching at the ADC}

	Recall that the abrupt depth change is met by traveling waves at dimensionless time $T_{ADC} = 0$, set to zero for convenience. The KdV equations describe wave dynamics over {\em long} timescales, physically $t \gg \timescale$, capturing weakly nonlinear and weakly dispersive effects. Meanwhile, evolution over shorter timescales is simply described by linear theory. The event of a wave crossing the ADC is precisely such a short-time event, and so we will employ the same matching conditions at the ADC that would result from linear theory. 

	More specifically, we assume continuity of the surface displacement, $\eta$, across the ADC \cite{whitham2011linear, rey1992propagation}. Since the propagation speed $c$ changes with depth, waves crossing the ADC must rapidly adjust in wavelength in order to match the oscillation frequency just upstream of the ADC \cite{whitham2011linear, rey1992propagation}. The normalized domain $x \in [-\pi,\pi]$ is scaled on the characteristic wavelength and so there is no change in the dimensionless wave-field $u(x,t)$, giving the condition
\begin{equation}
\label{detmatch}
u(x,t) \vert_{t=0^-} = u(x,t) \vert_{t=0^+}, 
\qquad \mbox{\em Deterministic matching condition}
\end{equation}	
We call \eqref{detmatch} the deterministic matching condition to contrast with its statistical counterpart introduced below. This matching condition is employed in the deterministic simulations of TKdV \eqref{TKdV}. We note that in the alternate formulation mentioned earlier, it is the product $\depth^{1/4} \eta$ that would match at the ADC \cite{johnson1997modern}.

	Now, from the perspective of statistical mechanics, consider the communication between the statistical ensembles, $\Gupdn$, upstream and downstream of the ADC. These two systems are in contact at the ADC, and so the upstream state with distribution $\Gup$ can be regarded as a thermal reservoir that influences the downstream distribution $\Gdn$. In fact, the above deterministic  condition directly leads to a simple description for the link. The quantity of interest is the outgoing Hamiltonian $\Hdn$, since its value just upstream of the ADC, $t=0^{-}$, is set by the incoming dynamics, and then this particular value is conserved thereafter in the outgoing dynamics. Since $u$ matches at the ADC, so must $\Hdn$, and, in fact, this matching holds for every individual trajectory. Recall that $\Hdn(t)$ is not conserved in the upstream dynamics and so its value varies for $t < 0$. However, appealing to weak Ergodicity, the ensemble mean $\meanup{\Hdn(t)}$ is expected to be independent of time. In particular, the value $\meanup{\Hdn \vert_{t=0^{-}} }$ is the same as the bare ensemble mean $\meanup{\Hdn}$. Afterwards, the particular value $\Hdn \vert_{t=0^{-}}$ is conserved in the downstream dynamics, on a trajectory-by-trajectory basis and for all $t>0$. Thus, the downstream measure $\Gdn$ must recover the same ensemble mean $\meanup{\Hdn \vert_{t=0^{-}} } = \meanup{\Hdn}$, producing the simple condition
\begin{align}
\label{statmatch}
\meanup{\Hdn} = \meandn{\Hdn}
\qquad \mbox{\em Statistical matching condition}
\end{align}

	This statistical matching condition, originally derived by \maetal \cite{majda2019}, imposes a relationship between the two inverse temperatures $\thup$ and $\thdn$. In particular, we view $\thup$ as given by the random-state of the incoming wave field. Thus, while we will treat $\thup$ as a parameter that can be varied to study various possible system states, the downstream $\thdn$ is determined directly by the matching condition \eqref{statmatch}, giving the functional dependence
\begin{equation}
\thdn = \transf \left( \thup \right)
\qquad \mbox{\em Transfer function}
\end{equation}
The {\em transfer function} $\transf$ will be a key link needed to relate the theory to experiments.

\subsection{Numerical computation of the transfer function}

	To compute the transfer function, we use a weighted random-sampling strategy. That is, we first sample the Fourier coefficients $\uhvec = \left\{ \uhat_{k}\right\}_{k=0}^{\Lambda}$ from a uniform distribution on the unit hypersphere $\En = 1$. The uniform sampling is achieved by first sampling from an isotropic Gaussian distribution and then normalizing to project onto the hypersphere \cite{abramov2003}. These samples are then weighted by the appropriate Gibbs measure \eqref{Gibbs} to compute the expectations needed in the statistical matching condition \eqref{statmatch}. That is, for an arbitrary quantity $Q$, the ensemble expectation corresponding to Hamiltonian $\Ham$ and inverse temperature $\theta$ is numerically approximated as
\begin{equation}
\label{weightmean}
\mean{Q}_{\theta} = \frac{\sumsamp Q_i \exp(-\theta \Ham_i)} {\sumsamp \exp(-\theta \Ham_i)}
\end{equation}
where $\Nsamp$ is the number of samples. Then, enforcing the statistical matching condition \eqref{statmatch} becomes a root-finding problem for the function
\begin{equation}
\label{rooteq}
\Fth(\thdn) =  \mean{\Hdn}_{\thdn} - \mean{\Hdn}_{\thup} = 0
\end{equation}
Recall that we consider $\thup$ as given, so that $\mean{\Hdn}_{\thup}$ is a constant that can be computed straightaway from \eqref{weightmean}. We then use the secant method with respect to the variable $\thdn$ to find a root of $\Fth(\thdn)$ to the desired tolerance.

For simply computing the transfer function, this weighted-sampling approach offers significant advantages over more sophisticated methods, such as such Markov Chain Monte Carlo (MCMC), in its ability to reuse the same samples of $\uhvec$ for several different values of $\thdn$. That is, we sample $\uhvec$ from the uniform distribution and compute the list of Hamiltonian values only once, then simply vary $\thdn$ in \eqref{rooteq} with the secant method until the root is found. An MCMC method, on the other hand, would require the sampling to restart from scratch for each value of $\thdn$, since the random steps in MCMC depend directly on the target distribution \eqref{Gibbs}. We will however, use MCMC to initialize direct numerical simulations of TKdV due to the superior efficiency for a single, given value of $\theta$.

\subsection{Deterministic simulations of TKdV}

We now detail the method for direct numerical simulation of the TKdV dynamical system \eqref{TKdV}. In particular, since the conservation of energy and Hamiltonian plays a central role in the emergent statistical features of the system, it is important for the numerical scheme to conserve these quantities over long time horizons. We therefore employ a symplectic integrator, which, by preserving oriented areas in phase space, conserves the energy and Hamiltonian exactly.

Rearranging the TKdV system \eqref{TKdV} gives
\begin{align}
\label{TKdV2}
&\pd{\uL}{t} = -  \frac{1}{2} C_3 \drat^{-3/2} \, \pd{}{x} \Proj (\uL)^2 - C_2 \drat^{1/2} \, \frac{\partial^3 \uL}{\partial x^3} = \RHS[\uL]
\end{align}	
where we have represented the right-hand side by the operator $\RHS[\uL]$. We employ a pseudo-spectral discretization of \eqref{TKdV2}, with de-aliasing applied to the quadratic nonlinear term $\Proj (\uL)^2$ according to the standard 2/3-rule. That is, we first pad $\uhat_k$ with zeros for $\Lambda < \abs{k} \le 3\Lambda/2$, transform to physical space and square to obtain $\uL^2$ on a fine grid, then transform back to spectral space and truncate to obtain the Fourier coefficients of $\Proj (\uL)^2$
For simplicity, we denote these Fourier coefficients $\usqdah_k$,
\begin{equation}
\Proj (\uL)^2 = \sum_{\abs{k} \le \Lambda} \usqdah_k e^{i k x}
\end{equation}
With this discretization, \eqref{TKdV2} can be recast in spectral space as a nonlinear ODE system
\begin{equation}
\label{TKdVDisc}
\td{}{t} \uhat_k =  -\frac{1}{2} C_3 \drat^{-3/2} \, ik \usqdah_k + C_2 \drat^{1/2} \, ik^{3} \uhat_k = \RHSh_k
\end{equation}
The quadratic nonlinearity represented by $\usqdah_k$ mixes the modes during evolution. We note the third-order linear term may become stiff for large $\Lambda$.

For time integration of \eqref{TKdVDisc}, we employ a 4th-order midpoint symplectic scheme \cite{mclachlan1993symplectic}. We introduce the spectral vectors $\uhvec = \left\{ \uhat_{k}\right\}_{k=0}^{\Lambda}$ and $\RHSvec = \left\{ \RHSh_{k} \right\}_{k=0}^{\Lambda}$, and let $\uhvec^n$ denote the solution at time $t_n$. The midpoint method has two intermediate stages and the auxiliary vectors $\mathbf{y}_{1},\mathbf{y}_{2}$:
\begin{align}
\label{SympStep1}
&\mathbf{y}_{1} - \uhvec^n = w_{1}\Delta t \, \RHSvec
\left[ \frac{1}{2}  \left(\mathbf{y}_{1} + \uhvec^n \right) \right] ,\\
&\mathbf{y}_{2}-\mathbf{y}_{1} = w_{2}\Delta t \, \RHSvec
\left[ \frac{1}{2} \left(\mathbf{y}_{2} + \mathbf{y}_{1}\right) \right] ,\\
\label{SympStep3}
&\uhvec^{n+1}-\mathbf{y}_{2} = w_{3}\Delta t \, \RHSvec
\left[ \frac{1}{2} \left( \uhvec^{n+1} + \mathbf{y}_{2}\right) \right] ,
\end{align}
with the time increments $w_{1}=\left(2+2^{1/3}+2^{-1/3}\right)/3$, $w_{2}=1-2w_{1}$, and $w_{3}=w_{1}$. The semi-implicit nature of \eqref{SympStep1}--\eqref{SympStep3} combined with the nonlinearity in $\RHSvec$ requires iteration. We split $\RHSvec$ into linear and nonlinear components, with the linear component being easily inverted since it is diagonal in spectral space. At each step of \eqref{SympStep1}--\eqref{SympStep3}, we perform Picard iteration on the nonlinear component until convergence is achieved with a tolerance of $\delta = 1\times10^{-10}$. The starting guess for the iterations is determined by quadratic extrapolation from the previous three stages.

	For the initial conditions, $\uhvec^0$, of the direct numerical simulation we sample from the upstream Gibbs ensemble $\Gup$ \eqref{Gibbs} with a prescribed inverse temperature $\thup$. We achieve this sampling via a Metropolis-Hasting Monte-Carlo algorithm as detailed in \cite{majda2019}.

\subsection{Scaling analysis to link inverse temperature to experiments}

	Since the inverse temperature, $\theta$, is a key modeling parameter, it would be highly desirable to normalize the system such that the value of $\theta$ does not depend sensitively on the truncation index as $\Lambda$ grows large. That way, $\theta$ can be interpreted as a real physical parameter that can be linked to the experiments and whose value does not depend sensitively on where one chooses to truncate the system. The one modeling parameter that is left to be set is $\lamfac$, which represents the number of characteristic wavelengths in the periodic domain. In what follows, we will determine reasonable constraints on $\lamfac$ that allow $\theta$ to be asymptotically independent of $\Lambda$.
	
	The invariant measure exhibits the proportionality $ \Gibbs_{\Lambda} \propto \exp(-\theta_{\Lambda} \Ham_{\Lambda})$, where we have made explicit the dependence of all quantities on $\Lambda$. In particular, if $\Ham_{\Lambda}$ were to depend sensitively on $\Lambda$ in expectation, then $\theta_{\Lambda}$ would need to compensate in order to produce the same invariant measure. Hence, it would be desirable to scale the system in such a way that $\Ham_{\Lambda}$ does not depend sensitively on $\Lambda$, at least in expectation. 
To achieve this insensitivity, we will appeal to the uniform measure $\Gz$ and the idea of equipartition of energy \cite{abramov2003}, since these two concepts afford simple scaling estimates.

Recall that $\Ham_{\Lambda}$ is composed of the cubic and quadratic components $\Hthree$ and $\Htwo$. Due to the odd symmetry of $\Hthree$, it is easy to see that $\meanz{\Hthree} = 0$ with respect to the uniform measure $\Gz$. The quadratic component, however, requires closer inspection. Due to Parseval's identity, $\Htwo$ can be written as
\begin{equation}
\Htwo = \frac{1}{2} \int_{-\pi}^{\pi} u_x^2 \dx = 2 \pi \sum_{k=1}^{\Lambda} k^2 \abs{\uhat_k}^2
\end{equation}
Due to the constraint $\En[\uL] = 1$, the equipartioned microstate is given by 
\begin{align}
&\abs{\uhat_k}^2 \approx \frac{1}{2 \pi \Lambda}	 \qquad \mbox{\em for equipartition of energy}
\end{align}
Then the expected value of $\Htwo$ under the uniform measure is
\begin{equation}
\label{HtwoExpect}
\mean{\Htwo}_0 = 2 \pi \sum_{k=1}^{\Lambda} {k}^ 2 \mean{\abs{\uhat_k}^2}_0 \sim \frac{1}{3} \Lambda^2
\end{equation}
where we have used the identity for the Gauss-like sum
\begin{equation}
\sum_{k=1}^{n} {k}^ 2 = \frac{1}{6} n(n+1)(2n+1) \approx \frac{1}{3} n^3
\end{equation}

	Importantly, \eqref{HtwoExpect} shows that the expected value of $\Htwo$ grows like $\Lambda^2$, which appears problematic for obtaining independence as $\Lambda \to \infty$. However, $\Htwo$ enters $\Ham$ in product with the coefficient $C_2$, which itself scales as $C_2 \sim \lamfac^{-2}$. Hence, obtaining the desired asymptotic independence with respect to $\Lambda$ simply requires that $\lamfac$ grow proportionally to $\Lambda$, as was already argued on physical grounds in Section \ref{tKdVSec}. 
Thus, any of the choices $\lamfac = \Lambda$, $\Lambda/2$, or $\Lambda/4$, discussed in that section would be valid. In particular, if $\lamfac = \Lambda$, then the characteristic wavelength in experiments corresponds to the smallest resolved wavelength in the dynamical system. It is perhaps more sensible to choose $\lamfac$ to be an intermediate value between 1 and $\Lambda$, so that some scales both larger than and smaller than the characteristic wavelength $\lam$ are resolved. As default, we will choose $\lamfac = \Lambda/2$, so that on a log-scale, the characteristic wavelength lies directly in the middle of the resolved wavelengths. 

	We note that the experiments discussed here preceded the development of the theory, and hence had no intent of mimicking a Gibbs measure in the wave forcing. We expect that if the experimental forcing were designed to mimic the Gibbs measure, in particular the precise spectral decay, then it would be more straightforward to assign a value to the model parameter $\lamfac$. We, however, leave that task for future research due to the significant cost of performing an entirely new set of experiments compared to the relative ease and great value in re-analyzing existing experimental data in light of the new theoretical developments.

\section{Comparison between theory and experiments}
\label{results}

With the experimental setup described and the theory outlined, we now present results comparing the two. Unless stated otherwise, all parameters used in the theory are taken directly from their experimental values listed in Table \ref{paramtable}. 

\subsection{Calibration of the inverse temperature}

At this point, all parameters appearing in the theoretical model have been linked directly to experimental parameters with the exception of the inverse temperature of the incoming flow, $\thup$.
Our strategy is to use the outgoing skewness as the main diagnostic to determine a realistic range for $\thup$. That is, for an input $\thup$, the downstream inverse temperature $\thdn$ is determined by \eqref{statmatch}, which ultimately sets the skewness of the outgoing wave-field.

\begin{figure}
\begin{center}
\includegraphics[width = 0.99 \linewidth]{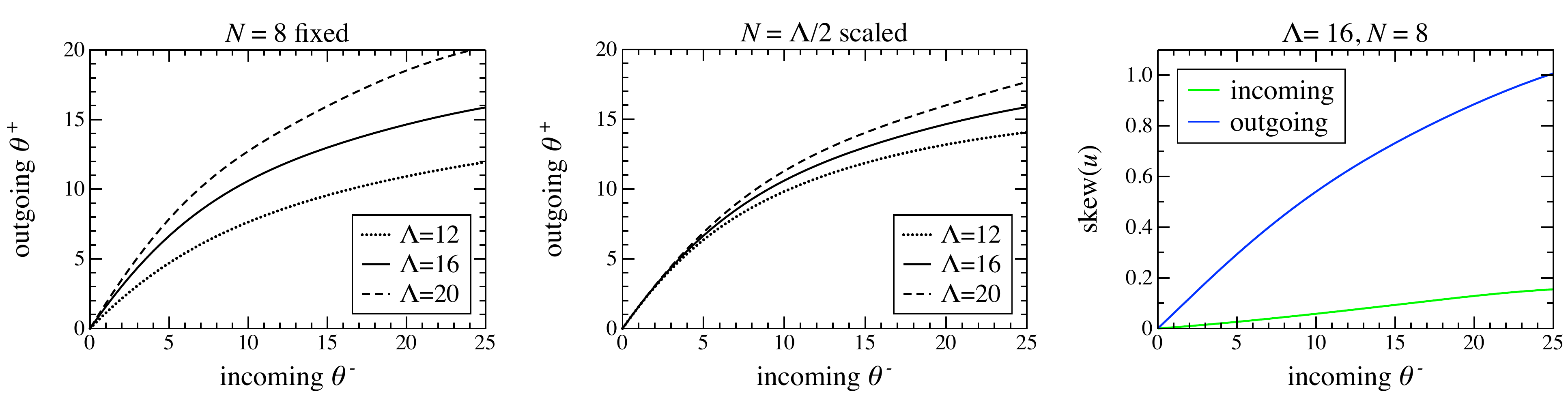}
\caption{
Analysis of the transfer function, $\thdn = \transf \left( \thup \right)$, and calibration of the inverse temperature. (a)--(b) The transfer function for three values of $\Lambda$ with either (a) $\lamfac = 8$ fixed or (b) $\lamfac = \Lambda/2$ scaled. Scaling $\lamfac$ with $\Lambda$ mitigates the sensitivity to $\Lambda$. (c) The incoming and outgoing skewness versus $\thup$ implied by the matching condition. In line with experiments, the skewness is enhanced significantly in the outgoing wave field.
}
\label{transfig}
\end{center}
\end{figure}
 
	Figures \ref{transfig}(a)-(b) show the transfer function $\thdn = \transf(\thup)$ that results from the statistical matching condition \eqref{statmatch}, for an incoming inverse temperature in the range $0 \le \thup \le 25$. Figure \ref{transfig}(a) shows the cases $\Lambda = $ 12, 16, and 20 with $\lamfac = 8$ fixed in each. In this figure, the transfer function changes significantly with $\Lambda$. Figure~\ref{transfig}(b) shows the same but with the scaling $\lamfac = \Lambda/2$ that was argued in the previous section. In this figure, the three curves come much closer to one another. Thus scaling $\lamfac$ appropriately greatly mitigates the sensitivity of the transfer function to $\Lambda$, though it does not completely remove the dependence.
 
	Next, Fig.~\ref{transfig}(b) shows the skewness of the incoming (green) and outgoing (blue) wave fields, as they depend on the incoming inverse temperature $\thup$ for the case $(\Lambda, \lamfac) = (16, 8)$. In line with experimental observations, the incoming skewness is small, while the outgoing wave skewness is much higher. Specifically, to capture the experimentally observed peak skewness range of 0.6-0.9 seen for the larger amplitudes in Fig.~\ref{ExpSpatialStats}(b), requires selecting $\thup$ in the range 10--25.

\subsection{Statistical comparison between theory and experiments}

	We now aim to compare the wave statistics measured in experiments against those that emerge from the TKdV theory. Throughout, we will focus on the representative set of experiments detailed in Table \ref{paramtable}, in which the peak downstream skewness was measured to be 0.83. 
With $(\Lambda, \lamfac) = (16, 8)$ fixed, Fig.~\ref{transfig}(c) indicates $\thup = 20$ as a reasonable value to attain the desired skewness. We thus run the deterministic TKdV simulations with these model parameters and with $\epsup$, $\delup$, and $\dratdn$ set to the representative values in Table \ref{paramtable}. We run the simulations to a sufficiently long dimensionless time of $t_{f} = 10$, with $dt = 5 \times 10^{-4}$, and with $10^3$ trajectories sampled from the upstream Gibbs measure $\Gup$.

\begin{figure}
\begin{center}
\includegraphics[width = 0.99 \linewidth]{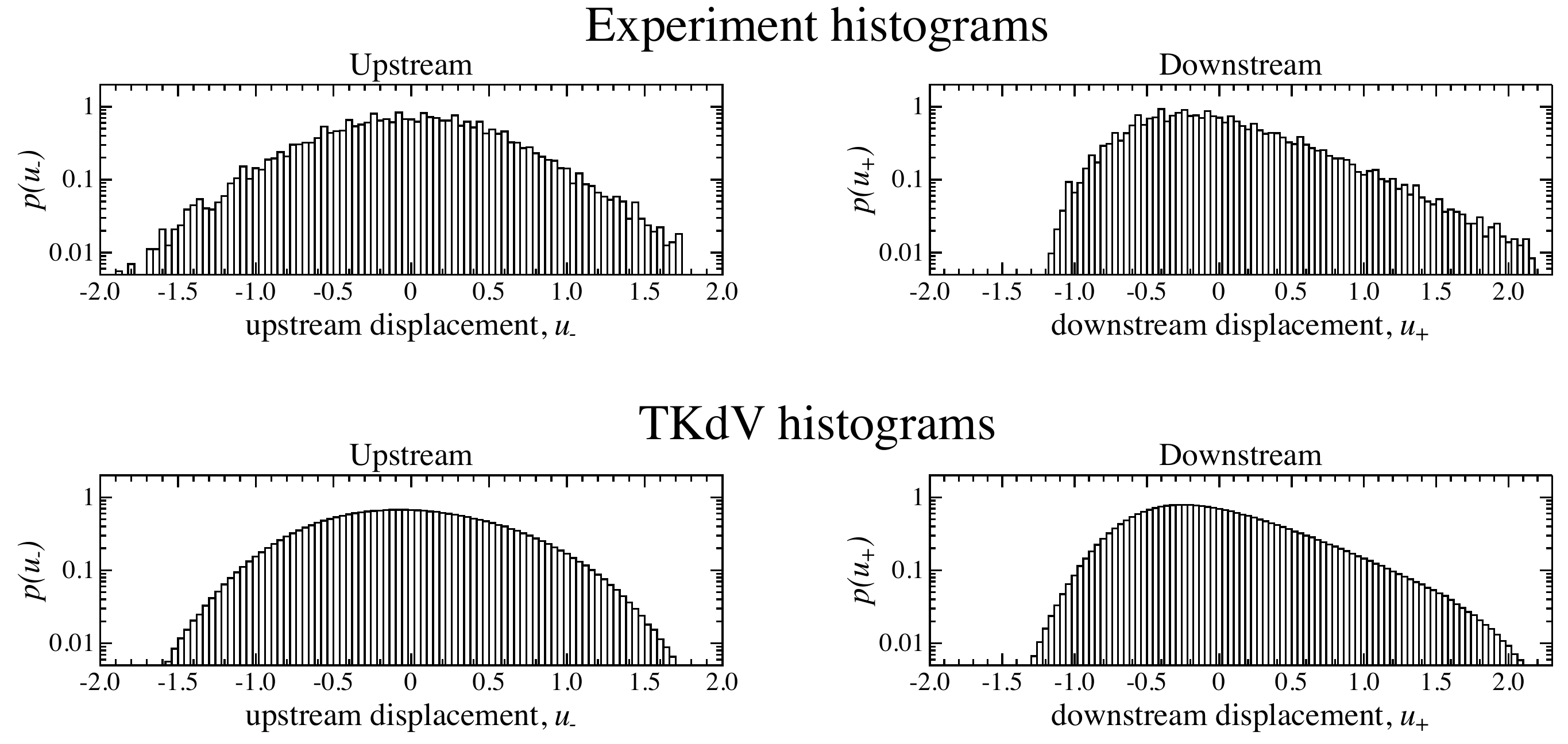}
\caption{ 
Comparison of displacement histograms from experiment and theory. Upon calibrating the inverse temperature $\thup$, the TKdV simulations recover the experimentally measured distributions in remarkable detail. In particular, they exhibit the transition from a nearly symmetric distribution upstream to a highly skewed distribution downstream.
}
\label{uhist}
\end{center}
\end{figure}
 
	In Fig.~\ref{uhist} we show the distributions of surface displacement that were measured in the experiments (top row) versus those that emerge from the TKdV simulations (bottom row). The experimental histograms represent exactly the same data depicted in Fig.~\ref{ExpDiagStats}, only converted to dimensionless displacement, $u$, for comparison against theory. Meanwhile, the TKdV distributions are extracted  from the long-time history of several trajectories sampled from the Gibbs ensemble, i.e.~mixed long-time/ensemble histograms. All histograms are shown on a semi-log scale to facilitate comparison of the tails, where extreme events lie.

	The comparison between experiments and theory in Fig.~\ref{uhist} is striking. As seen in the figure, both the experiments and theory show a transition from a nearly symmetric upstream distribution to a highly skewed distribution downstream. The TKdV theory not only captures this transition, but also recovers the shape of the resulting downstream distribution in remarkable detail. Nearly every feature that can be compared --- the decay rate of the tail, the position and value of the peak, the rapid cutoff for negative $u$ --- matches surprisingly well. This comparison offers compelling visual evidence for: (a) the predictive power of the TKdV framework, and (b) the successful calibration of the incoming inverse temperature.


	Next, we aim to make the same comparison for free-surface slopes. We note that the derivative, $\partial \eta / \partial x$, is already a dimensionless quantity with a simple physical interpretation, namely the slope of the free surface, whereas the interpretation of $\partial u/\partial \tilde{x}$ is tied to the characteristic values $\ampscale$ and $\lengthscale$.  
We therefore convert the theoretical calculated slopes values back to physical slopes via
\begin{equation}
\pd{\eta}{x} = 2 \pi^{3/2} \frac{\etastd}{\lamupdn} \pd{\uupdn}{\tilde{x}}
\end{equation}
So that the peak experimental wavelength $\lam$ (corresponding to $\freqp = 2$ Hz) would be mapped to the dominating TKdV spectral mode, $k = 1$,  this conversion formula has not been corrected with the extra factor of $\lamfac$ present in \eqref{scales}.
 
\begin{figure}
\begin{center}
\includegraphics[width = 0.99 \linewidth]{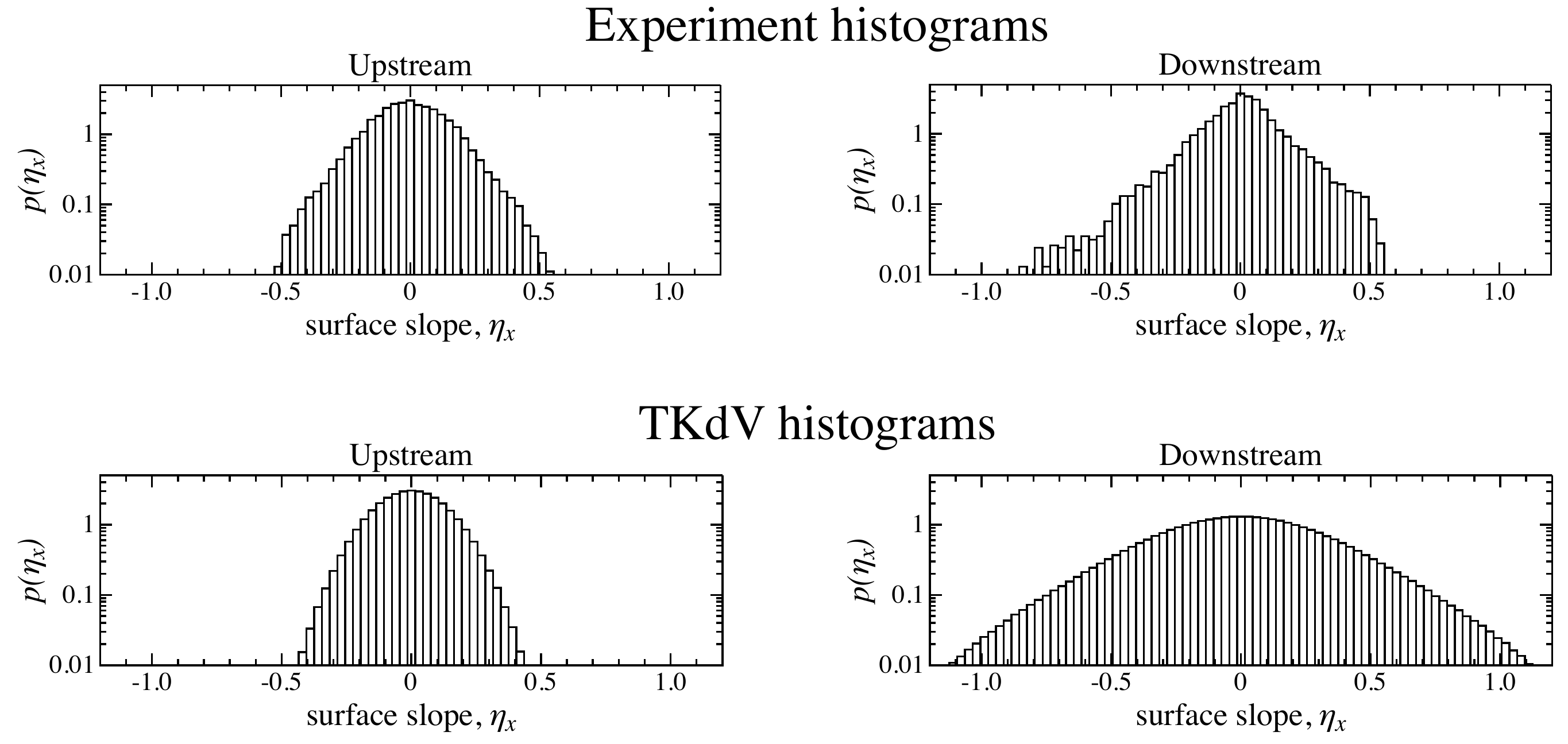}
\caption{
Comparison of surface slope histograms. The experiments and TKdV simulations show very similar slope distributions upstream. Downstream, the theory accounts for the spread of the distribution and the long exponential tails, while differences in the detailed shape of the distributions are also visible. Note the elevated uncertainty in the experimental data due to numerical differentiation of free-surface measurements.}
\label{slopehist}
\end{center}
\end{figure}
 
	Figure \ref{slopehist} shows the distributions of surface slope that result from experiments (top) and theory (bottom). These histograms show intriguing similarities and differences. First, the upstream slope distribution is captured well by the TKdV simulations, both in its nearly symmetric shape and its scale. We note that, while the standard deviation of displacement was input into the TKdV theory, the slope standard deviation was not. Downstream, both the experimental and theoretical slope distributions  spread significantly. Interestingly, in the experiments, it is not the standard deviation that increases significantly upon crossing the depth change ($\std(\eta_x)$ increases from 0.15 to 0.17), but instead the excess kurtosis, which jumps from a value of 0.46 upstream to 3.4 downstream. Likewise, in the TKdV theory, $\std(\eta_x)$ grows from 0.13 upstream to 0.33 downstream, and $\kurt(\eta_x)$ grows from -0.04 to 0.50. Thus, the theoretically predicted jump in kurtosis, though not as extreme as that measured in experiments, is quite significant. These elevated levels of kurtosis are associated with the distinct appearance of the downstream distributions, most notably the long, flat tails that appear in both experiments and theory.
 		
	Differences in the detailed shapes of these downstream distributions are also visible. First, we point out that the experimental measurements involve numerical differentiation of  surface displacement extracted from optical images, a process that unavoidably amplifies any noise that is present. We must therefore proceed with caution in comparing the experiment and theory, recognizing the possibility that observed discrepancies may be due to these measurement errors. Nonetheless, we notice that the theoretically predicted distributions remain symmetric downstream and the experimental ones skew towards negative slope. As mentioned in Section \ref{experiments}, negative skewness is consistent with a right-moving wave of steep leading surface (i.e.~negative slope). Additionally, the peak of the experimental distribution appears sharper than that of the theory.


\subsection{Wave dynamics and analysis of time scales}

\begin{figure}
\begin{center}
\includegraphics[width = 0.9 \linewidth]{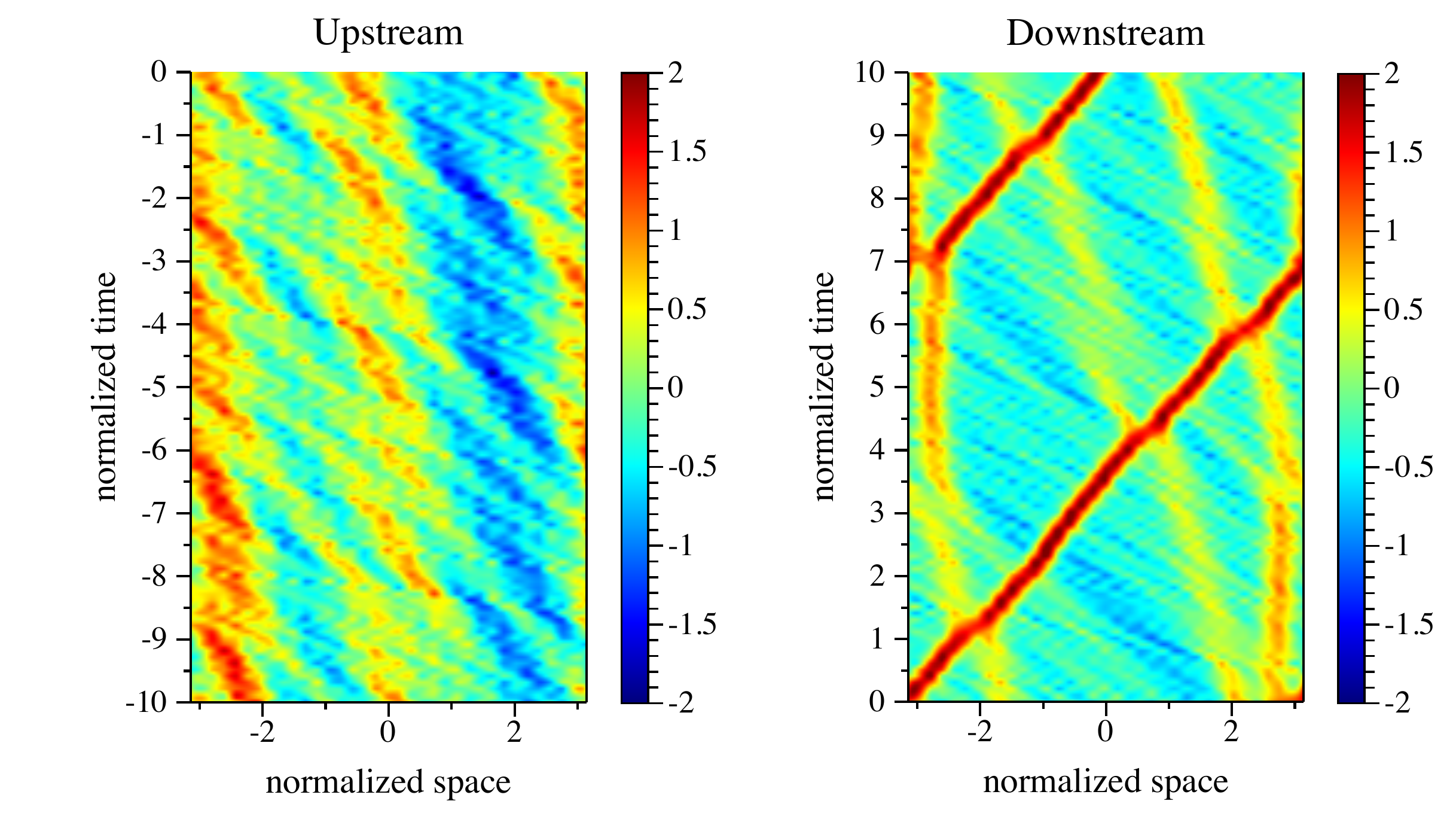}
\caption{
Sample upstream and downstream solution trajectories from the same ensemble TKdV simulations featured in Figs.~\ref{uhist} and \ref{slopehist}. The upstream solutions exhibit only leftward propagating waves and a high degree of regularity, while the downstream solutions exhibit both left and rightward moving waves along with more intermittency.
}
\label{trajectories}
\end{center}
\end{figure}

	As a complement to the above statistical comparison, we now examine the wave dynamics of a few individual trajectories from the TKdV simulations. Figure \ref{trajectories} shows example upstream and downstream solution trajectories from the same TKdV simulations that were used to produce the histograms in Figs.~\ref{uhist}--\ref{slopehist}. The dimensionless displacement, $u$, is represented by color in the domain $(x,t) \in [-\pi,\pi] \times [-10,10]$, where the ADC is encountered at $t=0$. 
	
	Visual differences between the upstream and downstream dynamics are apparent in Fig.~\ref{trajectories}. The upstream trajectory shows several waves of modest amplitude all propagating leftward. Here, the magnitude of the positive and negative displacements are comparable. This is in stark contrast to the downstream dynamics, which feature fewer waves of larger amplitude propagating in either direction. In particular, one large wave is seen to propagate from the lower-left corner of the figure to the upper right. A bias towards positive surface displacement is apparent in these downstream dynamics, without even viewing the histograms.

These observations can be rationalized with some simple observations regarding the structure of the TKdV system. Upstream of the ADC, since $\drat = 1$, the nonlinear effects are relatively weak and the dynamics are dominated by dispersion. Linearizing \eqref{TKdV} and introducing the ansatz $u_k = e^{i(kx - \omega_k t)}$ produces the dispersion relation for angular frequency $\omega_k = -C_2 k^3$, or for phase velocity $c_k = \omega_k/k = -C_2 k^2$ \cite{majdaqi2019}. In particular, the phase velocity is strictly negative, implying that waves only propagate leftwards, as is consistent with Fig.~\ref{trajectories}(a).

	Downstream of the ADC, however, the depth ratio changes to $\drat = 0.24$. This change significantly amplifies nonlinearity while suppressing dispersion, and thereby allows waves to propagate in either direction. In particular, the dispersionless limit of KdV is the Burgers-Hopf equation, for which the nonlinear advection is proportional in both amplitude and direction to the value of $u$. Hence, positive displacements would be expected to move rightwards, as is seen in Fig.~\ref{trajectories}(b).

	We remind the reader that the KdV framework tracks the long-time evolution of waves in a reference frame moving with the characteristic speed $c = \sqrt{g \depth}$ from linear theory. Hence, the above discussion of left-or-right going waves cannot be interpreted in the context of the experiments without an appropriate Galilean transformation. Thus, in the laboratory frame, all waves indeed propagate unidirectionally, from left to right, with a speed near $c$. The directions and speeds calculated by the TKdV framework simply quantify the deviation of the true wave speed from $c$.
		
	Finally, Fig.~\ref{trajectories} sheds some light on the timescales required for wave evolution in the TKdV framework. It appears that the normalized time of $t=10$ is roughly the correct timescale to observe substantial wave dynamics. More precisely, both the left-going waves in Fig.~\ref{trajectories}(a) and the right-going waves in Fig.~\ref{trajectories}(b) require a dimensionless time of about $t=7$ to traverse the entire periodic domain. For the representative experiments, the characteristic timescale, $\timescale$, is 2.8~sec upstream and 5.7~sec downstream. Hence, a dimensionless time of $t=7$ corresponds to roughly 20~sec and 40~sec upstream and downstream respectively. Given the wave speeds ($c = $ 110 cm/s upstream and 54 cm/s downstream), these timescales can be converted to distances traveled by the waves. In fact, a simple calculation using the definitions in Section \ref{theory} gives the distance traveled as $7 \lamfac g/(2 \pi \freqp^2)$, which, due to a serendipitous cancellation, is independent of depth. Hence, the distance traveled by the waves over a dimensionless time of $t=7$ is the same value upstream and downstream,  roughly 22 meters. Note that this distance is significantly greater than the 6-meter length of the experimental wave tank.

\begin{figure}
\begin{center}
\includegraphics[width = 0.99 \linewidth]{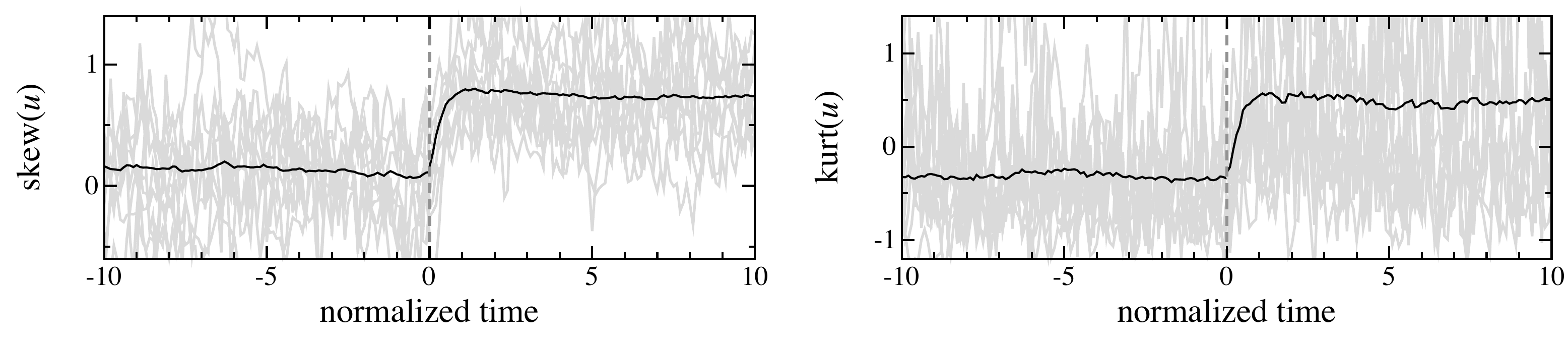}
\caption{
Dynamic evolution of surface-displacement skewness and kurtosis from the TKdV simulations. Faint gray curves show skewness and kurtosis of 100 individual trajectories, and the bold black curves show the corresponding ensemble mean. Skewness and kurtosis evolve over much shorter timescales than that required for waves to cross the domain.}
\label{skew-kurt}
\end{center}
\end{figure}
 
	These estimates raise an important question: if a distance of 22 meters is required for significant wave evolution under the KdV framework, how do the experiments exhibit substantial changes in wave statistics a much shorter distance downstream of the ADC? The results shown in Fig.~\ref{skew-kurt} help resolve this question. This figure shows the evolution of surface-displacement skewness (left) and kurtosis (right) computed by the same deterministic TKdV simulations as pictured in Fig.~\ref{trajectories}. The bold black curve shows the ensemble mean (ensemble size of 1000), while the faint gray curves show the skewness and kurtosis of 100 individual trajectories to give a sense for the variation involved. Importantly, the skewness and kurtosis evolve on a much shorter timescale than the aforementioned $t=7$. More precisely, skewness and kurtosis have already saturated to their asymptotic value by $t = 1$, and reach half of that value by $t = 0.18$. These dimensionless times correspond to travel distances of 310 cm and 56 cm respectively --- much shorter than the previously mentioned 22 meters, and on the same order as the relevant distances in the experiments. Thus, after crossing the ADC, the wave-field rapidly reconfigures itself enough to fundamentally alter its statistical distributions, and the timescale for this reconfiguration is much shorter than the time required for waves to cross the entire periodic domain.

\subsection{Explicit formula for outgoing skewness and experimental confirmation}

We now discuss arguably the most novel single result of the manuscript. In the recent theoretical study, \maetal derived an explicit formula for the outgoing wave-field skewness in terms of the system parameters \cite{majda2019}. More specifically, this formula relates the downstream skewness of surface displacement, $\skw(\eta)$, to the change in slope variance, $\var(\eta_x)$. Here, we recap the derivation of this formula and then test the prediction against new, direct experimental measurements.

The explicit formula for outgoing skewness arises directly from the statistical matching condition \eqref{statmatch}, which states that the downstream Hamiltonian must match in expected value at the ADC, i.e.~with respect to the incoming and outgoing Gibbs measures.
This condition can be written more explicitly as
\begin{equation}
\label{statmatchexp}
C_2 \dratdn^{1/2} \, \meanup{\Htwo} - C_3 \dratdn^{-3/2} \, \meanup{\Hthree } = 
C_2 \dratdn^{1/2} \, \meandn{\Htwo} - C_3 \dratdn^{-3/2} \, \meandn{\Hthree }
\end{equation}
Both the experiments and simulations show the upstream skewness to be negligible compared to its downstream counterpart, allowing the term with $\meanup{\Hthree }$ to be dropped in \eqref{statmatchexp}. With this approximation, \eqref{statmatchexp} yields the relationship
\begin{equation}
\label{H3H2ratio}
\frac{\meandn{\Hthree}} {\meandn{\Htwo} - \meanup{\Htwo}} = \frac{C_2}{C_3} \dratdn^2
\end{equation}

	Our next task is to convert this formula to physical variables so that it can be tested against experimental data. First, following the definition of $\Hthree$ in \eqref{H3H2} and the definition, $u = \eta/(\pi^{1/2} \etastd)$, a straightforward calculation gives
\begin{align}
\label{H3convert}
\meandn{\Hthree} = \frac{\pi}{3} \meandn{u^3} = 
\frac{1}{3 \pi^{1/2}} \frac{\meandn{\eta^3}}{ \etastd^3} = 
\frac{1}{3 \pi^{1/2}} \skwdn(\eta)
\end{align}
This equation expresses a direct relationship between the expected value of $\Hthree$ and the outgoing skewness of surface displacement. Next, we must convert the $\Htwo$-quantities, which involve surface slope. 
Using the definition of $\Htwo$ in \eqref{H3H2} and $\lengthscale_+$ in \eqref{scales} gives
\begin{align}
\label{H2convert}
&\meanupdn{\Htwo} = \pi \meanupdn{ u_x^2} = 
\left( \frac{\lamfac \lamdn}{2 \pi \etastd} \right)^2 \varupdn(\eta_x)
\end{align}
We note that the downstream scale $\lengthscale_+$ is used in this conversion, since it is the downstream Hamiltonian that is matched in \eqref{statmatch}. This formula links the expected value of $\Htwo$ to the variance of surface slope.

Inserting \eqref{H3convert} and \eqref{H2convert} into \eqref{H3H2ratio}, and using the definitions of $C_2$ and $C_3$ from \eqref{C3C2}, gives the remarkably explicit formula
\begin{equation}
\label{SkewPred}
\frac{\skwdn(\eta)} {\vardn(\eta_x) - \varup(\eta_x)} = \frac{1}{3} \epsup^{-3} \dratdn^3 
\end{equation}
This formula links the skewness of the outgoing wave field to the change in the variance of surface slope. In particular, it predicts their ratio to scale as the {\em inverse cube of wave amplitude}, $\epsup^{-3}$, and the {\em cube of the depth ratio}, $\dratdn^3$. It is a rather unexpected relationship, as it would have been difficult to anticipate that the displacement skewness and slope variance, among all possible variable combinations, are so intimately related.
 
\begin{figure}
\begin{center}
\includegraphics[width = 0.99 \linewidth]{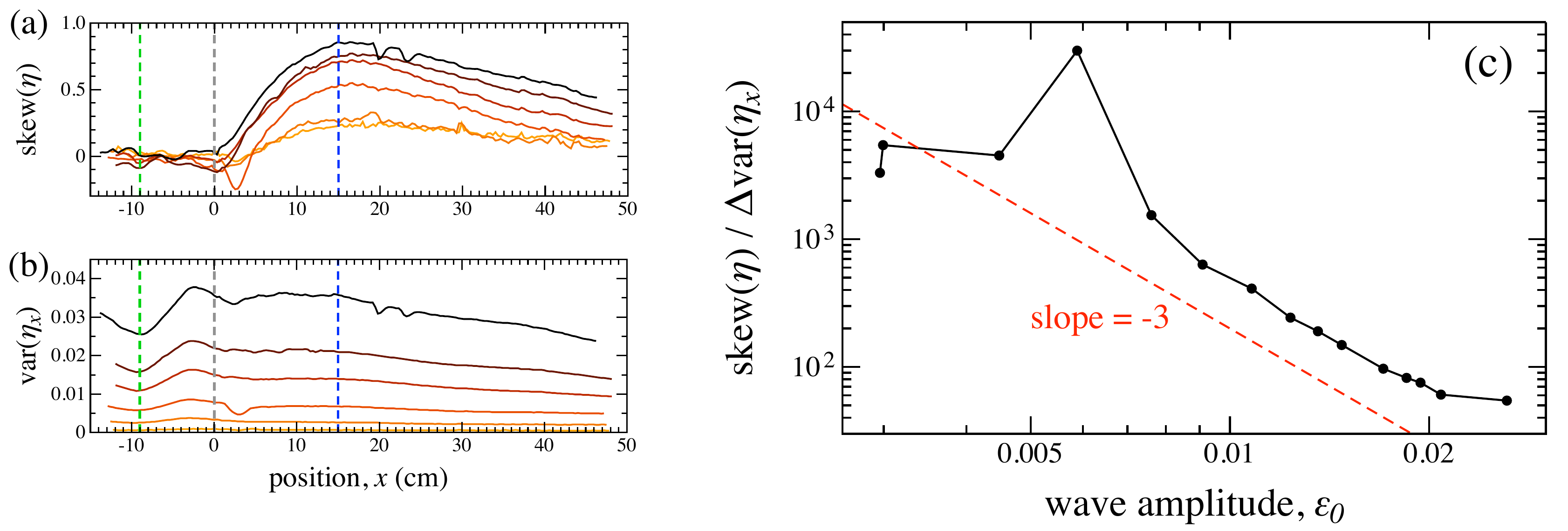}
\caption{
Experimental confirmation of the explicit formula for outgoing skewness \eqref{SkewPred}.
(a)--(b) Spatial variation of $\skw(\eta)$ and $\var(\eta_x)$ for several different amplitudes. The maximum of $\skw(\eta)$ and the minimum of $\var(\eta_x)$ each occur at a location that is insensitive to amplitude, $x = $15 cm (blue) and -9 cm (green) respectively.
(c) Measurements of the ratio $\skw(\eta) / \Delta \var(\eta_x)$ from 15 different experiments plotted against the dimensionless wave amplitude $\epsup$. For sufficiently large amplitude, the measurements follow the predicted $\epsup^{-3}$ power law closely.
}
\label{SkewRat}
\end{center}
\end{figure}
 
	Fortunately, it is a relationship that can be tested directly against the experimental measurements, specifically the type of data that was presented in Fig.~\ref{ExpSpatialStats}. Formula \eqref{SkewPred} draws focus to the spatial variation of displacement skewness, which we replot for convenience in Fig.~\ref{SkewRat}(a), and the variance of surface slope, shown in Fig.~\ref{SkewRat}(b). To test \eqref{SkewPred}, we must select a representative downstream position to evaluate $\skwdn(\eta)$ and $\vardn(\eta_x)$, and a representative upstream position for $\varup(\eta_x)$. In short, we select the same representative locations  used to evaluate the histograms in Figs.~\ref{ExpDiagStats}, \ref{uhist}, and \ref{slopehist}, namely $x = 15$ cm downstream and $x = -9$ cm upstream. These positions are indicated by the blue and green vertical dashed lines in Fig.~\ref{SkewRat}(a)--(b). Recall that $x = 15$ cm corresponds to the peak of the downstream skewness and kurtosis. Meanwhile, we see in Fig.~\ref{SkewRat}(b) that $x = -9$ cm corresponds to a slight dip in $\var(\eta_x)$.

	With these locations selected, we evaluate the ratio $\skw(\eta) / \Delta \var(\eta_x)$ for each experiment and plot the result against dimensionless wave amplitude, $\epsup = \etastd / \dup$ on a log-log scale in Fig.~\ref{SkewRat}(c). While the data shows significant variation at small amplitudes, it shows a well defined decreasing trend at larger amplitudes. In particular, the red dashed line shows the power-law $\epsup^{-3}$ predicted by \eqref{SkewPred}, which captures the experimental trend remarkably well.

\section{Concluding remarks}
\label{conclusion}

	This manuscript extends the parallel experimental and modeling efforts of \boetal \cite{bolles2019} and \maetal \cite{majda2019} concerning the emergence of anomalous wave statistics from abrupt changes in bottom topography. The theoretical framework is based on deterministic and statistical analysis of the TKdV system, with exploitation of the Hamiltonian structure and the associated invariant Gibbs measures. The theory depends crucially on matching the incoming and outgoing invariant measures at the abrupt depth change, so as to link the incoming and outgoing states. Throughout, we have emphasized the synergy between the experiments and theory, with experimental data informing theoretical advancements and model predictions motivating new experimental measurements.
	
	Careful calibration of the inverse temperature against the experimental data allowed for a detailed comparison between experimental and theoretically predicted wave statistics. The outgoing distributions of surface displacement predicted by the TKdV framework capture in remarkable detail the experimental measurements. We have extended this statistical analysis to surface slope, a line of inquiry motivated by the importance of $\Htwo$, the slope standard deviation, in the theory. The comparison of slope statistics shows intriguing similarities, while also some quantitative differences in the shape and skewness of the outgoing distributions.
	
	Finally, \maetal \cite{majda2019} derived an explicit formula for the skewness of the outgoing wave field, which we have tested against direct experimental measurements. Specifically, the formula predicts how the ratio of displacement skewness to slope variance, $\skw(\eta) / \Delta \var(\eta_x)$, depends on wave amplitude. The experimental measurements conclusively confirm the inverse-cube dependence on wave amplitude, thus highlighting the predictive power of the TKdV statistical mechanics framework.
	
	We note that since the ratio, $\skw(\eta) / \Delta \var(\eta_x)$, exhibits such a clear dependence on input parameters, it could prove a useful diagnostic for anomalous wave observations in the ocean. In particular, this quantity seems to provide a signature for anomalous waves that are triggered by abrupt depth changes. Examination of this ratio in field data is an exciting avenue for future research.

\section*{Acknowledgements}
CTB acknowledges support from the IDEA grant at Florida State University, as well as from the Geophysical Fluid Dynamics Institute. 
MNJM acknowledges support from the Simons Foundation, award 524259. 
This research of A.J.M. is partially supported by the Office of Naval Research N00014-19-S-B001. 
D.Q. is supported as a postdoctoral fellow on the grant.

\bibliographystyle{plain}
\bibliography{wavesbib}

\end{document}